\documentclass[journal]{IEEEtran}
\usepackage{cite}
\usepackage{graphicx}
\usepackage{amsmath}
\usepackage{siunitx}
\usepackage{makecell}
\graphicspath{{../pdf/}{../jpeg/}}

\newif\ifarxiv
\arxivtrue

\begin{document}
%
\title{Helium-Gas-Cooled Cryogenic Current Comparator Integrated with a Quantum Resistance Standard}
%
%

\author{Yuma~Okazaki,
		Takehiko~Oe,~
		and~Nobu-Hisa~Kaneko
\thanks{Y.~Okazaki, T.~Oe, and N.-H.~Kaneko are with National Institute of Advanced Industrial Science and Technology (AIST), National Metrology Institute of Japan (NMIJ), Tsukuba 305-8568, Japan e-mail: yuma.okazaki@aist.go.jp.}
\thanks{This research was partly supported by JSPS KAKENHI Grant No. JP23H01861, and the CSTI BRIDGE Program “Development of Wide-Range Current Measurement Technology with Assured Quantum Traceability” (Funding Agency: QST).}
}

%
%

%



\maketitle

\begin{abstract}
We report a cryogen-free cryogenic current comparator (CCC) system for precision resistance measurements. A helium-gas chamber was developed and installed on the 4 K stage of a cryogen-free dilution refrigerator equipped with a pulse-tube cryocooler. The CCC probe was housed in this chamber and was cooled through helium gas serving as a heat exchange medium. The metrological performance of this helium-gas-cooled CCC was evaluated through precision resistance-ratio measurements and found to be comparable to that obtained under liquid-helium cooling. A quantum Hall resistance (QHR) device was also integrated into the same refrigerator, enabling QHR/100~$\Omega$ resistance-ratio measurements. The type-A uncertainty reached the 1~n$\Omega$/$\Omega$ level within an averaging time of 100~s, and the resulting resistance-ratio measurements agreed well with conventional liquid-helium-based measurements at the level of a few n$\Omega$/$\Omega$. The system provides this level of measurement performance while consuming less than 1~L of helium gas per thermal cycle.
\end{abstract}


\IEEEpeerreviewmaketitle

\section{Introduction}
%
%
%
%

\IEEEPARstart{A}{} cryogenic current comparator (CCC) is an indispensable instrument in electrical metrology\cite{2009DRUNG, 2021POIRIER}, including resistance calibrations based on the quantum Hall resistance (QHR) at a precision of a few $10^{-9}$~$\Omega$/$\Omega$ \cite{2013SCHOPFER}, fundamental studies of next-generation quantum resistance standards\cite{2022OKAZAKI, 2024PATEL, 2025RODENBACH}, and ultra-low-current measurements\cite{2015DRUNG, 2020GIBLIN, 2023KANEKO}. A CCC consists of a set of windings, a superconducting quantum interference device (SQUID) magnetometer for detecting current imbalance, and multiple magnetic shields. 
Most CCC probes are operated in liquid-helium cryostats. The recent increase in the cost and the limited availability of liquid helium\cite{2025Hu} have therefore motivated efforts to reduce helium consumption while maintaining metrological performance\cite{2012LAWSON, 2018Wang, 2015JANSSEN, 2019RIGOSI, 2025CHAE, 2026TAUPIN}. 

Recent advances in pulse-tube cryocoolers have made sub-4 K temperatures readily accessible without consuming liquid helium. Their practical application to CCCs is not straightforward because pulse-tube cryocoolers inherently generate mechanical vibrations, which may introduce excess noise and possibly degrade precision measurements\cite{ 2026TAUPIN, 2004TOMARU, 2016Mykkanen,2016Kalra,2017Olivieri}.
In addition, implementation of conventional CCC probes in cryogen-free refrigerators presents a thermal-engineering challenge. Cryogen-free refrigerators rely on solid thermal conduction inside a vacuum-insulated environment. However, CCC probes originally designed for liquid-helium immersion contain multiple nested structures for magnetic shielding, making efficient cooling of the internal components difficult. Helium gas is an attractive heat-exchange medium because it remains gaseous even at 4.2~K below its saturation pressure (about 100~kPa) and provides efficient cooling of the internal components that are otherwise difficult to thermally anchor. Despite these advantages, the implementation of helium-gas-cooled CCCs and their achievable metrological performance remain unexplored.

In this work, we develop a helium-gas chamber for operating a commercially available CCC in a cryogen-free dilution refrigerator. Non-circulating helium gas introduced into this chamber serves as a heat-exchange medium to cool the CCC probe that is originally designed for operation in liquid helium. The pressure dependence, stability, and metrological performance of the helium-gas-cooled CCC are investigated through SQUID characterization and precision resistance-ratio measurements. Furthermore, we integrate a QHR device and a CCC within a single refrigerator despite the inherently conflicting magnetic environments required by the two systems: a magnetic field of approximately 10~T for the QHR and $10^{-15}$~T level magnetic sensitivity for the CCC. The successful operation of the CCC under the magnetic conditions required for QHR measurements was demonstrated through winding-ratio error tests and precision QHR/100~$\Omega$ resistance-ratio measurements.

\section{Experiments}
\subsection{System description}

\begin{figure*}[tb]
 \centering
\includegraphics[width=\textwidth]{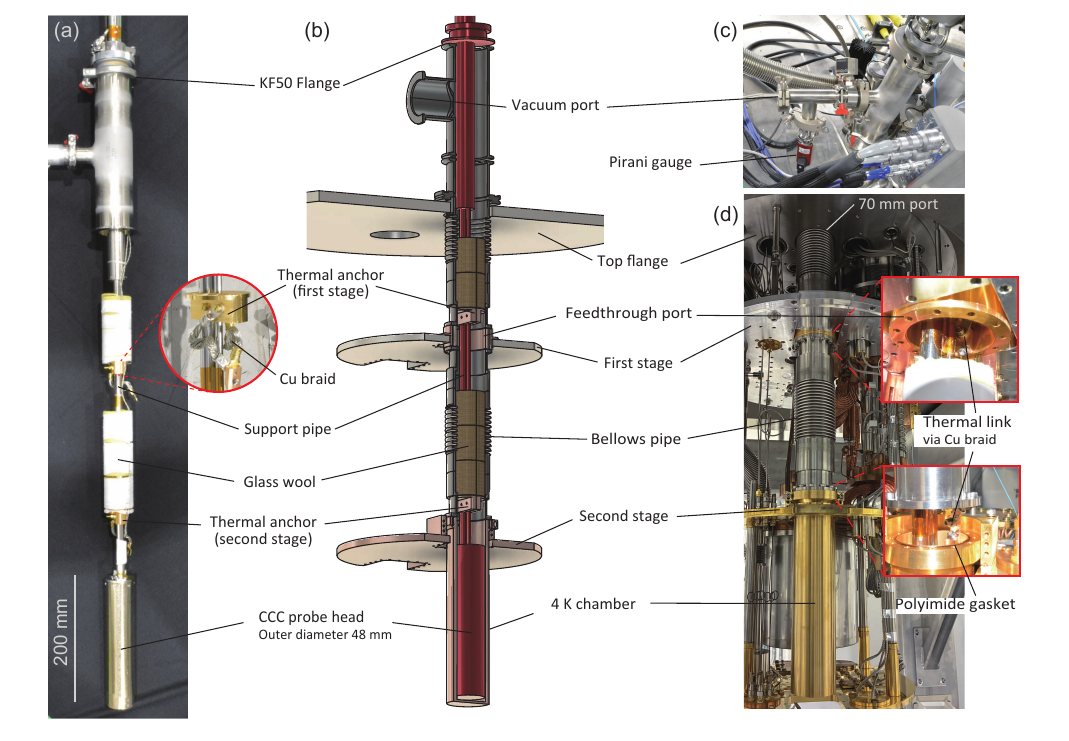}
 \caption[]{The developed helium-gas-cooled CCC. (a) Photograph of the CCC probe. The circled inset shows an enlarged view of the thermal anchor. (b) Cutaway schematic of the helium-gas chamber installed in a cryogen-free refrigerator. (c) Exterior view of the assembled gas chamber. (d) Interior view of the gas chamber installed inside the refrigerator. The upper and lower red-boxed insets show partially assembled views of the first-stage and second-stage thermal anchor structures, respectively.}
 \label{fig01}
\end{figure*}

In this study, we used a commercially available 12-bit CCC originally developed at the Physikalisch-Technische Bundesanstalt \cite{2009DRUNG, 2009GOTZ, 2013DRUNG_BCU} and now widely adopted by national metrology institutes worldwide for resistance-ratio measurements\cite{2020NIM, 2020KRISS,2025PTB, 2026CEM}.
The CCC is equipped with a binary winding array ranging from 1 to 2048 turns and six additional windings, allowing arbitrary turn ratios within a 12-bit range to be configured.
The CCC system also includes high-performance electronics comprising an ultra-low-noise chopper amplifier for null detection, a compensation network for resistance bridge operation, and SQUID electronics for highly sensitive flux detection. This system enables resistance-ratio measurements up to 1~M$\Omega$/QHR within practical measurement times.
This CCC probe was originally designed for operation with the probe head immersed in liquid helium.
To operate this CCC in a cryogen-free refrigerator, we utilized the first and second stages of a dilution refrigerator (BlueFors LD250) equipped with a pulse-tube cryocooler, providing nominal cooling powers of 45~W at 45~K (at the first stage) and 1.35~W at 4.2~K (second stage), respectively.
To effectively cool the CCC probe head inside the vacuum insulation chamber of the refrigerator, a helium-gas chamber was developed and installed through a 70~mm-diameter line-of-sight port, as shown in Fig.~1.
The CCC probe consists of a cylindrical probe head with a diameter of 48~mm, a stainless-steel support pipe, and a KF50 vacuum flange as shown in Fig.~1(a).
The developed helium gas chamber consists of a cylindrical 4~K chamber thermally anchored to the second stage, a feedthrough port mounted on the first stage, a bellows pipe connecting the two stages, and associated vacuum components for gas handling.
The 4~K  chamber and feedthrough port were fabricated from oxygen-free copper and gold-plated without a nickel underlayer.
The bellows section is formed from 316 stainless steel and has an inner diameter of 49.5~mm, an outer diameter of 65~mm, and a wall thickness of 0.15~mm.

The CCC probe was inserted vertically through the top flange such that the probe head was fully accommodated inside the 4~K chamber. 
Oxygen-free copper thermal anchor clamps were attached to the support pipe at positions corresponding to the first and second stages of the refrigerator, as shown in Fig.~1(a). After insertion of the CCC probe, the thermal anchors were connected to blind tapped holes on the inner wall of the feedthrough port and the 4~K chamber via oxygen-free copper braids with a cross-sectional area of 8~mm$^2$ and a length of 80~mm, as shown in the inset of Fig.~1(d). These thermal anchors are essential to suppress the heat flow through the support pipe. The heat load associated with the helium-gas-cooled CCC and the detailed design of the components are described in Supplementary Material.
A 0.125~mm-thick polyimide film cut into an O-ring shape was used as a sealing gasket to hermetically seal all joints in the cryogenic section. The absence of detectable helium leakage from the gas chamber into the vacuum insulation chamber was confirmed using a leak detector after the system reached its base temperature.
The free volume between the refrigerator stages inside the gas chamber may lead to unwanted heat transfer due to helium gas. To suppress such heat conduction, the bellows section was filled with a cylindrical glass-wool insulator, as shown in Fig.~1(b). The insulators were fabricated from a glass-wool board by cutting into a cylindrical shape with a diameter of 48~mm and a height of 50~mm. Five glass-wool pieces were attached along the support pipe using PTFE tape prior to insertion, as shown in Fig.~1(a) and inserted into the gas chamber together with the CCC probe.

 The vacuum port was connected to both a pumping system for evacuation and a helium gas cylinder (99.995~\% purity) for gas introduction. The helium gas flow was manually regulated using precision needle valves with a control range of 0.1~L/min.
The chamber pressure was monitored using a Pirani gauge (TPR 270, Pfeiffer Vacuum) below 1~kPa and a mechanical compound pressure gauge above 1~kPa, because the Pirani gauge is not sufficiently accurate for helium gas at higher pressures. The Pirani gauge has a calibration factor of 1.0 for helium gas below 100~Pa; therefore, no correction was applied in this pressure range.

\subsection{Helium gas pressure dependence}

\begin{figure}[tb]
 \centering
\includegraphics[width=\columnwidth]{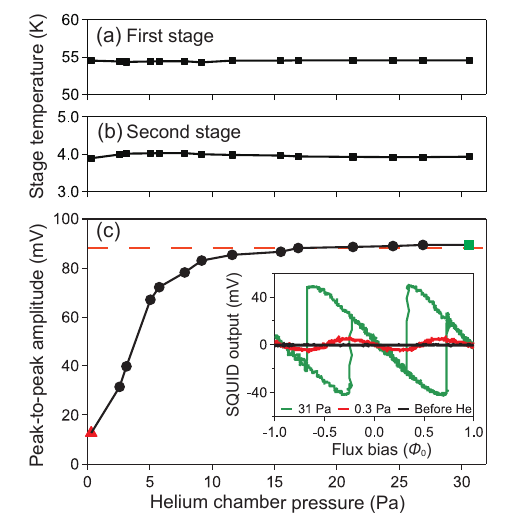}
 \caption[]{Helium gas pressure dependence of the system performance. (a, b) Temperatures of the first stage (a) and second stage (b) as a function of the helium gas pressure in the chamber. (c) Peak-to-peak amplitude of the SQUID $\phi$-V curve as a function of helium gas pressure. The dashed line indicates the typical $\phi$-V amplitude of 88~mV obtained for the same CCC under liquid-helium cooling. The red triangle and green square correspond to the measurements at 0.3~Pa and 31~Pa, respectively, whose $\phi$-V curves are shown in the inset. Inset: amplified SQUID output voltage as a function of flux bias measured before helium-gas introduction (Before He), and after helium-gas introduction at chamber pressures of 0.3~Pa (red) and 31~Pa (green), where $\Phi_0$ is the magnetic flux quantum.}
 \label{fig02}
\end{figure}

Prior to cooldown, the helium gas chamber was evacuated using a turbomolecular pump. Pumping was continued until the refrigerator reached its base temperature. The operation of the CCC was first monitored through the $\phi$-V characteristics of the integrated SQUID \cite{2009DRUNG} during the helium-gas introduction process.
The inset of Fig.~2(c) shows oscilloscope X-Y traces of the amplified SQUID output voltage, with the DC offset subtracted, under flux modulation. Before helium gas was introduced, the SQUID output showed no observable response to the flux modulation, as indicated by the black trace in the inset. This result indicates that the SQUID remained above its superconducting transition temperature despite the second-stage temperature being below 4.2~K, suggesting insufficient cooling of the internal components of the CCC in the absence of helium gas.
After helium-gas introduction, the SQUID exhibited the characteristic $\phi$-V curve at a chamber pressure of approximately 0.3~Pa (red trace in the inset). The larger amplitude observed at 31~Pa (green trace in the inset) indicates improved SQUID performance with increasing chamber pressure. Since a larger $\phi$-V amplitude provides higher flux sensitivity, the peak-to-peak amplitude was investigated as a function of chamber pressure, as shown in Fig.~2(c). Note that the measurements were conducted after waiting at least 10~min following helium gas introduction to allow the system to reach the steady-state condition. The $\phi$-V amplitude increased with increasing chamber pressure and eventually saturated at pressures above approximately 20~Pa. The saturated amplitude was comparable to the value obtained when the same CCC is operated in liquid helium, approximately 88~mV as indicated by the dashed line. This result confirms that sufficient cooling of the internal components can be achieved in the presence of helium gas. Notably, less than 1~L of room-temperature helium gas was sufficient to achieve a chamber pressure of 31~Pa.

We also monitored the temperatures of the first and second stages as a function of helium gas pressure, as shown in Figs.~2(a) and (b). No significant temperature change was observed at either stage, indicating that heat transfer through the helium gas was sufficiently suppressed. 
At helium gas pressures of several tens of kPa, however, a measurable increase in stage temperature was observed. Nevertheless, the developed chamber maintained its integrity up to 100~kPa without detectable helium leakage. 

Although CCC operation was possible immediately after helium gas introduction, increased measurement fluctuations were observed before thermal equilibrium was established. Therefore, all precision measurements were performed after waiting at least one hour following helium gas introduction.

\subsection{$10\,\mathrm{k\Omega}/100\,\Omega$ Resistance-Ratio Measurements}

\begin{figure}[tb]
 \centering
\includegraphics[]{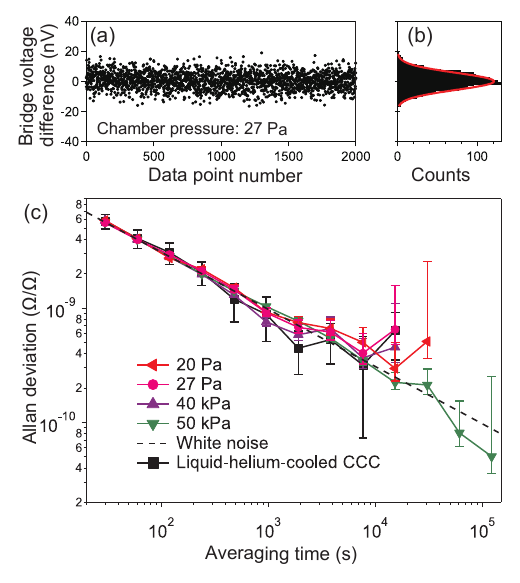}
 \caption[]{Measurements of 10~k$\Omega$/100~$\Omega$ resistance ratio using the 4000:40 winding configuration. (a) Results of 2000 consecutive measurements of the bridge voltage difference plotted as a function of measurement index at a helium gas pressure of 27~Pa. (b) Histogram of the measurement results shown in (a) together with a Gaussian fit. (c) Allan deviation of the measured resistance ratio. The colored symbols represent measurements performed with the CCC operated in the helium-gas chamber at pressures of 20~Pa, 27~Pa, 40~kPa, and 50~kPa. The black squares represent reference data obtained using the same CCC cooled by liquid helium. Error bars indicate the 68.3~\% confidence intervals obtained from a chi-square analysis. The dashed line shows the white-noise dependence ($\propto \tau^{-1/2}$).}
 \label{fig03}
\end{figure}

We then demonstrate 10~k$\Omega$/100~$\Omega$ resistance-ratio measurements using the helium-gas-cooled CCC. Two standard resistors, SR104 (10~k$\Omega$) and SR102 (100~$\Omega$), housed in a temperature-controlled air bath maintained at 23.00~$^\circ$C, were used in this study. Measurement currents of \SI{50}{\micro\ampere} and 5~mA were applied to the 10~k$\Omega$ and 100~$\Omega$ resistors, respectively. The CCC was operated in a standard 4000:40 winding configuration. Further details of the CCC operation and bridge configuration can be found in Refs.~\cite{2009DRUNG, 2013DRUNG_BCU}.

Figure~3(a) shows the results of 2000 consecutive measurements of the bridge voltage difference determined from measurements obtained with opposite current polarities. One measurement cycle required 30~s in total, consisting of 15~s at each current polarity, including about 1~s for current reversal. For each polarity, the data acquired in the last 5~s were used in the calculation. Consequently, the 2000 consecutive measurements took approximately 16.7~h. Despite such a long continuous measurement, the results are randomly distributed around the mean value and show no observable drift or discontinuities. The histogram shown in Fig.~3(b) confirms that the fluctuations follow a Gaussian distribution.
To elucidate their statistical profile, the Allan deviation of the resistance ratio was calculated, as shown in Fig.~3(c). Measurements were performed at helium gas pressures of 20~Pa, 27~Pa, 40~kPa, and 50~kPa. Results obtained using the same CCC operated in liquid helium are also shown for comparison. 
All data sets show good agreement and follow the white-noise dependence up to averaging times approaching $10^4$~s. The corresponding type-A uncertainty reaches the 1~n$\Omega$/$\Omega$ level within an averaging time of 1000~s. This uncertainty is comparable to the typical uncertainties reported for similar resistance-ratio measurements in international comparisons using CCCs\cite{2020NIM, 2020KRISS, 2025PTB ,2026CEM, 2020BIPM_EMK12_NMIJ}. These results indicate that the developed system provides performance suitable for precision resistance metrology despite its liquid-helium-free operation.

In addition to operation in the low-pressure range, measurements at relatively high pressures of 40~kPa and 50~kPa were performed. Although the high-pressure measurements exhibited an Allan deviation comparable to that obtained at low pressure, an increase in the stage temperature was observed.
The high-pressure measurements were conducted primarily to evaluate the performance limits of the developed helium-gas-cooled CCC. Because of the increased thermal load and the higher helium consumption, operation in the kPa range is not considered practical. All subsequent measurements were therefore performed in the low pressure range.

\subsection{CCC operations under QHR measurements}

\begin{figure}[tb]
 \centering
\includegraphics[width=\columnwidth]{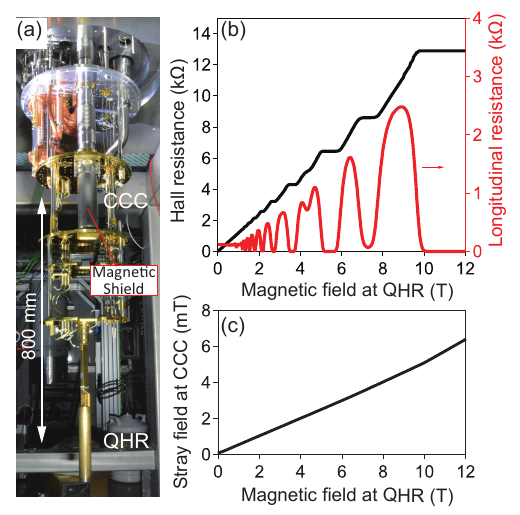}
 \caption[]{Integration of a CCC and a quantum Hall resistance (QHR) device within a single refrigerator. (a) Photograph of the refrigerator setup. The QHR device is mounted at the field center of a superconducting magnet, approximately 800~mm below the CCC probe. The gas chamber was covered with a magnetic shield. (b) Hall resistance and longitudinal resistance of the QHR device as a function of magnetic field, showing the development of the quantum Hall state. (c) Stray magnetic field measured near the position of the CCC probe head as a function of the magnetic field applied to the QHR device.}
 \label{fig04}
\end{figure}

We then investigated the integration of a QHR device and a CCC within the same refrigerator and demonstrated resistance-ratio measurements traceable to the QHR.
Figure~4(a) shows the refrigerator setup incorporating both the CCC probe with the helium-gas chamber and the QHR device.  The bottom plate corresponds to the mixing chamber stage of the dilution refrigerator, which provides a base temperature of approximately 20~mK. A GaAs-based QHR device was mounted at the field center of a 12~T superconducting magnet.
The measured Hall and longitudinal resistances as a function of magnetic field are shown in Fig.~4(b). Clear signatures of the quantum Hall effect were observed, namely quantization of the Hall resistance accompanied by vanishing longitudinal resistance. In particular, the $\nu = 2$ integer quantum Hall plateau used as the primary resistance standard was observed near 11~T.
In the present setup, the QHR device was located approximately 800~mm below the helium-gas chamber. The stray magnetic field generated by the superconducting magnet may affect the CCC performance because of the extremely high magnetic field sensitivity of the integrated SQUID. To evaluate this effect, the stray magnetic field was measured  using a Hall-effect magnetometer placed near the CCC position. This magnetometer was placed outside the vacuum insulation chamber at room temperature, at the same height as the CCC probe head and approximately 100~mm outside the CCC position, and was oriented to measure the vertical component of the magnetic field.
Figure~4(c) shows the measured stray magnetic field as a function of the magnetic field applied to the QHR device. At the operating field of 11~T corresponding to the center of the $\nu = 2$ plateau, the stray field remained below 7~mT.
To enhance the magnetic shielding, the helium-gas chamber was covered with an additional magnetic shield closed at one end with a wall thickness of 1.5~mm, as shown in Fig.~4(a). Indeed, no measurable change in the SQUID flux-bias point was observed before and after the application of the magnetic field, indicating that the stray magnetic field was sufficiently shielded.

Prior to precision resistance-ratio measurements based on the QHR, a ratio error test was performed under this magnetic-field condition, i.e.~11~T, to evaluate the performance of the CCC in the presence of the stray magnetic field. Ratio errors were measured for the winding combinations listed in Table~\ref{tab1}. These measurements were performed at a chamber pressure of 20~Pa. The uncertainties in the table were calculated as the standard deviations of $N=20$ repeated measurements divided by $\sqrt{N-1}$.

All winding combinations exhibit normalized ratio errors at or below the $10^{-9}$ level, indicating no measurable degradation of winding-ratio consistency even in the presence of the stray magnetic field. Furthermore, the ratio errors for typical winding configurations, such as 4001:31 and 4000:40, can be calculated from the results in Table~\ref{tab1}.
The estimated ratio error of the 4001:31 winding configuration, which is used for the QHR/100~$\Omega$ resistance-ratio measurements is
\begin{equation}
\frac{\Delta (N_{4001}/N_{31})}{4001/31} = (0.065 \pm 0.123) \times 10^{-9},
\end{equation}
and that of the 4000:40 winding configuration is
\begin{equation}
\frac{\Delta (N_{4000}/N_{40})}{4000/40} = (-0.076 \pm 0.068) \times 10^{-9},
\end{equation}
where the uncertainties were obtained by propagating the uncertainties listed in Table~\ref{tab1}. The estimated ratio errors are at the level of $10^{-10}$, making their contribution to the overall uncertainty negligible, because the combined standard uncertainty for these measurements is typically a few n$\Omega$/$\Omega$ as reported in Refs.~\cite{2020NIM, 2020KRISS, 2025PTB, 2026CEM, 2020BIPM_EMK12_NMIJ}.
These results confirm that no measurable winding-ratio error was observed even in the presence of the stray magnetic field generated during QHR operation, thereby enabling the integration of the CCC and QHR systems within a single refrigerator.

\begin{table}[tb]
\centering
\caption{Ratio error tests under magnetic-field condition.}\label{tab1}
\begin{tabular}{rl S[table-format=1.3] @{\(\;\;\pm\)} S[table-format=1.3]}
\hline
$N$ & Winding combination under test & \multicolumn{2}{c}{\rule{0pt}{2.8ex}$\Delta N/N\;(\times 10^{-9})$} \\
\hline
1 & 1a-1b                                    &  0.295 & 1.182 \\
2 & 1a+1b-2a                                 & -0.914 & 1.165 \\
2 & 1a+1b-2b                                 &  0.438 & 1.237 \\
4 & 2a+2b-4a                                 & -0.560 & 0.432 \\
4 & 2a+2b-4b                                 &  0.226 & 0.485 \\
8 & 4a+4b-8                                  & -0.062 & 0.286 \\
16 & 4a+4b+8-16a                              &  0.040 & 0.144 \\
16 & 4a+4b+8-16b                              & -0.069 & 0.120 \\
17 & 16a+1a-17                                &  0.182 & 0.139 \\
32 & 16a+16b-32                               &  0.057 & 0.023 \\
64 & 16a+16b+32-64                            &  0.026 & 0.023 \\
128 & 16a+16b+32+64-128                        &  0.005 & 0.021 \\
256 & 16a+16b+32+64+128-256                    &  0.006 & 0.010 \\
512 & 16a+16b+32+64+128+256-512a               &  0.008 & 0.004 \\
512 & 16a+16b+32+64+128+256-512b               &  0.012 & 0.005 \\
1024 & 512a+512b-1024                           &  0.001 & 0.006 \\
2048 & 512a+512b+1024-2048                      &  0.007 & 0.005 \\
2048 & \makecell[l]{ 1a+1b+2a+4a+8+16a+32+64+128\\ $\,\,\,\,\,\,$+256+512a+1024-2048 }
                                          &  0.010 & 0.021 \\
\hline
\end{tabular}
\end{table}

\subsection{QHR/100~$\Omega$ resistance-ratio measurements}

To demonstrate QHR/100~$\Omega$ resistance-ratio measurements, the CCC was operated using a 4001:31 winding configuration. Measurement currents of \SI{38.74}{\micro\ampere} and 5~mA were applied to the QHR and 100~$\Omega$ resistor, respectively. Other settings for the CCC measurements such as averaging time were identical to those used for the 10~k$\Omega$/100~$\Omega$ measurements.
Figures~5(a)-(c) show the results of the QHR/100~$\Omega$ comparison. Figure~5(a) presents 1000 consecutive measurements of the bridge voltage difference obtained at a helium gas pressure of 20~Pa, while Fig.~5(b) shows the corresponding histogram together with a Gaussian fit. Figure~5(c) shows the Allan deviation of the measured resistance ratio as a function of averaging time. For comparison, results obtained using the same CCC cooled by liquid helium are also shown. 

Several notable differences can be observed between the Allan deviation shown in Fig.~5(c) and that obtained for the 10~k$\Omega$/100~$\Omega$ comparison in Fig.~3(c). In both cases, the Allan deviation follows the expected white-noise dependence up to nearly $10^4$~s as indicated by the dashed lines. However, the overall noise level is lower for the QHR/100~$\Omega$ comparison, that is, the 10~k$\Omega$/100~$\Omega$ comparison required 1000~s of averaging to reach the 1~n$\Omega$/$\Omega$ level, whereas the QHR/100~$\Omega$ comparison reached the same uncertainty level within 100~s.
Another difference is the dependence on the CCC cooling method. For the 10~k$\Omega$/100~$\Omega$ comparison, the results obtained under helium-gas cooling and those under liquid-helium cooling agree with each other. In contrast, the Allan deviation obtained under liquid-helium cooling for the QHR/100~$\Omega$ comparison was approximately 35~\% lower than that obtained with helium-gas cooling.
This difference can be understood by considering the dominant noise sources in each measurement. In the 10~k$\Omega$/100~$\Omega$ comparison, Johnson noise from the 10~k$\Omega$ resistor at 300~K dominates the measurement uncertainty. Consequently, the influence of the CCC cooling method was negligible compared with the dominant Johnson noise contribution and could not be resolved within the measurement uncertainty. In contrast, for the QHR/100~$\Omega$ comparison, the QHR device was operated at a temperature below 1~K, resulting in substantially reduced thermal noise. Under these conditions, noise originating from the CCC system itself becomes a more significant contribution to the overall uncertainty, allowing differences associated with the CCC cooling method to be resolved.
The origin of the observed noise difference between helium-gas cooling and liquid-helium cooling remains unclear. Appendix~A presents an evaluation of pulse-tube-related noise, which may contribute to the observed difference, although no definitive conclusion can be drawn from the present data.

\begin{figure}[tb]
 \centering
\includegraphics[width=\columnwidth]{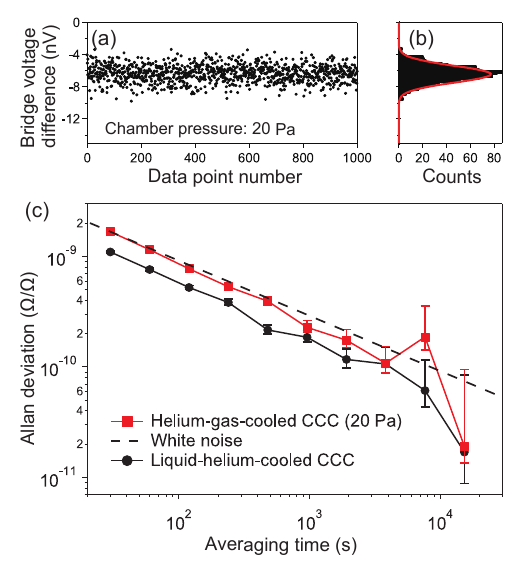}
  \caption[]{Measurements of QHR/100~$\Omega$ resistance ratio using the 4001:31 winding configuration. (a) Results of 1000 consecutive measurements of the bridge voltage difference plotted as a function of data point number at a helium gas pressure of 20~Pa. (b) Histogram of the measurement results shown in (a) together with a Gaussian fit. (c) Allan deviation of the measured QHR/100~$\Omega$ resistance ratio. The squares and circles represent data obtained with the CCC operated under helium-gas cooling and liquid-helium cooling, respectively. Error bars indicate the 68.3~\% confidence intervals obtained from a chi-square analysis. The dashed line shows the white-noise dependence ($\propto \tau^{-1/2}$).}
 \label{fig05}
\end{figure}

Despite this moderate increase in noise, the helium-gas-cooled CCC maintained a white-noise behavior and achieved a type-A uncertainty below 1~n$\Omega$/$\Omega$ within 100~s. This uncertainty is comparable to the typical uncertainties reported in international comparisons for this measurement using CCC bridges \cite{2020NIM,2020KRISS,2025PTB,2026CEM,2020BIPM_EMK12_NMIJ}, confirming the suitability of the present system for quantum resistance standards. These results demonstrate that a helium-gas-cooled CCC can be integrated with a QHR system in a single refrigerator while maintaining the measurement performance required for precision resistance metrology.

\section{Discussion}
We performed QHR/100~$\Omega$ resistance-ratio measurements using a total of seven standard resistors, including one SR102 and six HRU-101 resistors. The HRU-101 resistors were manufactured and commercialized by Alpha Electronics Corp{.} in collaboration with NMIJ\cite{2011SAKAMOTO, 2012KANEKO}. The set of resistors serves as the reference set for routine calibration services, and each resistor is regularly calibrated against the QHR standard at least twice a year. The long-term stability of all resistors has been characterized for more than a decade, providing well-established reference values for evaluating the performance of the developed system.
All resistance-ratio measurements were consistent with historical results obtained from QHR/100~$\Omega$ resistance-ratio measurements using the CCC operated under liquid-helium cooling. The mean deviation from the resistance values predicted by linear fits to the historical data of the seven resistors was $0.77$~n$\Omega$/$\Omega$ with a standard deviation of $1.38$~n$\Omega$/$\Omega$. The maximum observed deviation was 2.7~n$\Omega$/$\Omega$. This excellent agreement, together with the detailed investigations presented in this paper, demonstrates that the developed helium-gas-cooled CCC system provides performance suitable for precision resistance metrology.

By employing helium gas as a heat-exchange medium, we successfully operated a CCC originally designed for liquid-helium cooling on the 4~K stage of a cryogen-free refrigerator. At a helium gas pressure as low as 20~Pa, the CCC achieved performance comparable to that obtained under conventional liquid-helium operation, as verified by both the SQUID $\phi$-V characteristics and the Allan deviation of resistance-ratio measurements.
The developed system was also integrated with a QHR device within the same cryostat. Even in the presence of the strong magnetic field required for the QHR, the CCC successfully passed winding-ratio error tests, enabling the proper operation of the CCC even in the magnetic environment required for QHR measurements. Moreover, the QHR/100~$\Omega$ resistance-ratio results were consistent with those previously obtained using the same CCC operated under liquid-helium cooling.
Although a slow drift and short-period fluctuations were observed in the refrigerator stage temperature, their effects were effectively canceled by the offset subtraction associated with current reversal and did not measurably affect the resistance-ratio measurements (see Supplementary Material).
The CCC performance was reproducible over multiple thermal cycles. The refrigerator can be continuously operated for more than one month without detectable helium leakage from the gas chamber, and the system has accumulated more than one year of successful operational experience as of the time of writing.
When operated in the practical low-pressure range below 100~Pa, the system consumed less than 1~L of helium gas, which corresponds to a liquid-equivalent volume of 1~mL.
Our cryogenic technique is valuable for maintaining resistance standards under increasingly constrained liquid-helium supply. Furthermore, the successful integration of the CCC and QHR systems within a single cryostat reduces experimental complexity and laboratory footprint requirements while preserving metrological performance.

These results confirm that the developed system satisfies the performance requirements for primary resistance standard calibrations. In fact, the system has been used for resistance calibration services at NMIJ since September 2025.

\appendices


\section{Noise Properties under Pulse-tube Operation}

\begin{figure}[tb]
 \centering
\includegraphics[width=\columnwidth]{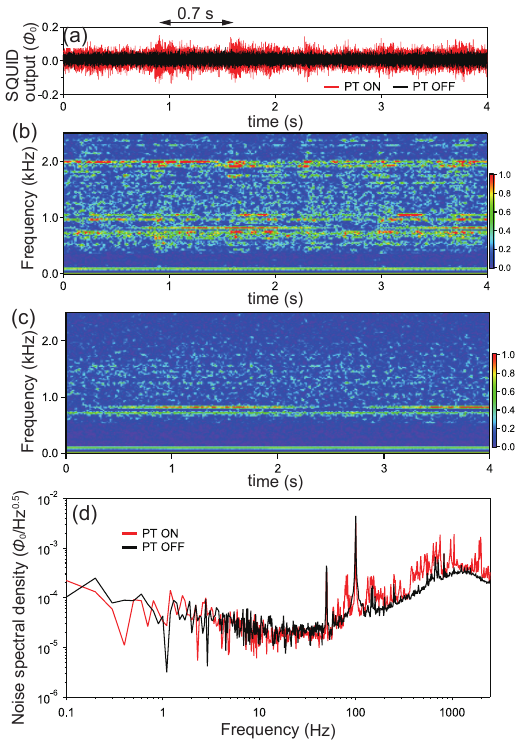}
 \caption[]{(a) Time-domain SQUID output measured with the pulse-tube refrigerator operating (PT ON, red) and temporarily stopped (PT OFF, black). (b, c) Short-time Fourier transform spectrograms of the SQUID output for the PT ON and PT OFF conditions, respectively. The same color scale, expressed in units of m$\Phi_0/\mathrm{Hz}^{0.5}$, is used in both spectrograms. (d) Noise spectral density obtained from the Fourier transform of the SQUID output for the PT ON (red) and PT OFF (black) conditions.}
 \label{fig06}
\end{figure}

Pulse-tube cryocoolers generate mechanical vibrations and acoustic noise, which can introduce unwanted electrical noise\cite{2026TAUPIN, 2016Mykkanen, 2016Kalra, 2017Olivieri}. In this appendix, we evaluate their influence on the developed CCC system.

The CCC was configured for 10~k$\Omega$/100~$\Omega$ resistance-ratio measurements using the same 4000:40 winding configuration as that used in the experiments presented in Fig.~3. The chamber pressure was 20~Pa. The time-domain SQUID output voltage was recorded using a digital oscilloscope, as shown in Fig.~\ref{fig06}(a). To highlight the influence of pulse-tube operation, the same measurement was also performed with the pulse tube temporarily stopped (PT OFF). The PT OFF data were acquired immediately after stopping the pulse tube and before the stage temperatures began to rise. The SQUID output voltage was converted to the corresponding magnetic flux in units of the magnetic flux quantum, $\Phi_0$.
Figure~\ref{fig06}(a) shows that the SQUID output exhibits a higher noise level during pulse-tube operation. In addition, a periodic noise with a period of approximately 0.7~s is visible, corresponding to the cycle of the pulse tube.
To investigate the frequency-domain characteristics of the noise, short-time Fourier transform spectrograms for the PT ON and OFF are shown in Figs.~\ref{fig06}(b) and (c), respectively, while the corresponding noise power spectral densities are plotted in Fig.~\ref{fig06}(d). Figure~\ref{fig06}(d) shows no significant difference between the PT ON and OFF below approximately 80~Hz. In contrast, a clear increase in noise is observed under pulse-tube operation at higher frequencies. This feature is also evident in the spectrograms shown in Figs.~\ref{fig06}(b) and (c).

Because CCC resistance-ratio measurements are fundamentally DC measurements, low-frequency noise is expected to be the dominant contribution to measurement uncertainty. From this viewpoint, the spectra shown in Fig.~\ref{fig06}(d) alone would suggest only a limited impact on precision resistance-ratio measurements. However, it is known that high-frequency noise can be rectified through SQUID nonlinearities and appear as an excess contribution in the low-frequency measurement result\cite{2015DRUNG}.
It is therefore conceivable that the high-frequency noise associated with pulse-tube operation may influence precision DC resistance-ratio measurements through similar mechanisms. At present, however, no direct evidence is available to determine whether pulse-tube-induced noise was responsible for the difference observed between helium-gas cooling and liquid-helium cooling in Fig.~5(c). Clarifying the relationship between pulse-tube-induced noise and resistance-ratio measurement performance remains a subject for future investigation.

\ifCLASSOPTIONcaptionsoff
  \newpage
\fi

\vfill


\end{document}


\title{Supplementary Material for\\
Helium-Gas-Cooled Cryogenic Current Comparator Integrated with the Quantum Resistance Standard}

\author{Yuma~Okazaki, Takehiko~Oe,  and~Nobu-Hisa~Kaneko}

\maketitle

\setcounter{figure}{0}
\renewcommand{\thefigure}{S\arabic{figure}}
\renewcommand{\theequation}{S\arabic{equation}}
\renewcommand{\thetable}{S\arabic{table}}
\def\bibsection{\section*{\refname}} \def\bibsection{\section*{\refname}} 

\begin{abstract}
This Supplementary Material provides estimates of the thermal loads associated with the bellows and CCC support pipe. It also presents additional details of the design, fabrication, and assembly of the helium gas chamber for readers interested in developing similar systems. Furthermore, the stability of the refrigerator stage temperature and its influence on the CCC measurements are evaluated.
\end{abstract}

\section{Thermal design of the system}
Installing the helium gas chamber and the CCC probe into a cryogen-free refrigerator introduces additional heat loads that may affect the cooling performance of the refrigerator. Therefore, an order-of-magnitude thermal analysis was carried out prior to fabrication.

The heat flow through the structure was estimated using the one-dimensional steady-state heat conduction equation. The heat flux is given by
\begin{equation}
J(x)
=
-Ak(T(x))
\frac{\partial T(x)}{\partial x},
\label{eq:heat_flux}
\end{equation}
and the steady-state condition requires
\begin{equation}
\frac{\partial J(x)}{\partial x}
=
0,
\label{eq:heat_conservation}
\end{equation}
where $A$ denotes the cross-sectional area and $k(T)$ is the temperature-dependent thermal conductivity. Thermal conductivity data were taken from the NIST Cryogenic Material Properties Database \cite{NISTDB}. The equations were numerically solved using the finite-difference method, from which the heat load at each stage was estimated.

The first heat-conduction path considered was the bellows section connecting the stages. In the simulation, the temperatures of the top plate, first stage, and second stage were assumed to be 300~K, 55~K, and 4.2~K, respectively. The bellows was modeled as a stainless-steel structure with a wall thickness of 0.15~mm, an effective cross-sectional area of 27~mm$^2$, and an unfolded path length of 320~mm. The resulting heat loads were estimated to be 0.24~W and 0.014~W at the first and second stages, respectively.

\begin{figure}[tb]
\centering
\includegraphics[]{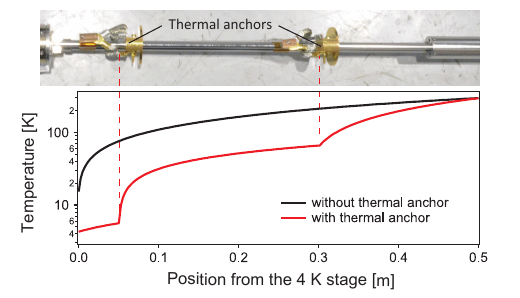}
\caption[]{Plot of the calculated temperature distribution along the CCC support pipe. The red and black traces represent the cases with and without intermediate thermal anchors, respectively. The photograph shows the CCC probe with thermal anchors installed. The dashed lines indicate the positions of the thermal anchors corresponding to the first and second stages.}
\label{figS01}
\end{figure}

Next, heat conduction through the CCC support pipe was evaluated. The support pipe was modeled as a stainless-steel rod with a cross-sectional area of approximately 60~mm$^2$. Figure~\ref{figS01} shows the calculated temperature distributions with and without thermal anchors at the first and second stages. The positions of the thermal anchors were 0.05~m and 0.30~m from the 4~K stage, respectively, as indicated by the vertical dashed lines. The thermal anchor was assumed to be connected to the stage through a copper braid with a cross-sectional area of 8~mm$^2$ and a length of 80~mm. Without thermal anchoring, elevated temperatures extend toward the probe head, resulting in an estimated heat load of 0.35~W at the second stage. In contrast, thermal anchoring at the intermediate stages effectively suppresses heat propagation, reducing the second-stage heat load to 0.054~W. The calculated temperature distributions clearly demonstrate the importance of thermal anchoring. Consistent with this analysis, no SQUID signature was observed experimentally in the absence of thermal anchors, even when helium gas pressures as high as 100~kPa were introduced.
\begin{table}[tb]
\centering
\caption{Estimated heat loads at first and second stages.} \label{tabS1}
\begin{tabular}{rll}
\hline
 & First stage & Second stage \\
\hline
Bellows & 0.24~W &  0.014~W \\
CCC pipe w/ thermal anchor & 0.75 W &  0.054 W \\
CCC pipe w/o thermal anchor & - & (0.35~W) \\
Total w/ thermal anchor & $\sim 1$~W & 0.068 W\\
\hline
Cooling power & 45~W at 45~K & 1.35~W at 4.2~K\\
\hline
\end{tabular}
\end{table}

The estimated heat loads are summarized in Table~\ref{tabS1}. Heat conduction through the CCC support pipe exceeds that through the bellows, making thermal anchoring essential for suppressing heat flow into the probe head. Nevertheless, the total heat load introduced by the CCC probe and the helium gas chamber remains small compared with the nominal cooling power available at each refrigerator stage.
The present analysis does not include heat transfer through the helium gas. Although quantitative modeling of heat transport through helium gas is beyond the scope of this work, no measurable increase in stage temperature was observed in the low-pressure operating regime, as shown in Figs.~2(a) and 2(b). This result suggests that heat transfer through the helium gas appears to be sufficiently suppressed by the glass-wool insulation.

\section{Design and Fabrication of the Gas Chamber}

\begin{figure}[t]
 \centering
\includegraphics[]{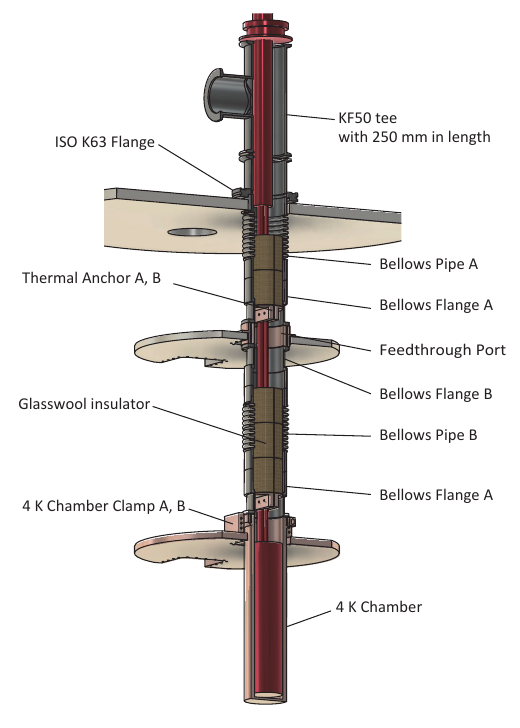}
 \caption[]{Cutaway schematic model of the helium gas chamber and the name of the components.}
 \label{figS02}
\end{figure}

The low-temperature section of the gas chamber was designed to accommodate the 48~mm-diameter probe head while maintaining an outer diameter below 70~mm, allowing installation through the 70~mm line-of-sight port.
The component names are defined in Fig.~\ref{figS02}, and their materials and fabrication methods are summarized in Table~\ref{tab:parts}. All components were designed using three-dimensional parametric CAD (Autodesk Fusion), and two-dimensional technical drawings were generated from the CAD models. The designed components were custom fabricated by specialized machining companies and vacuum-component manufacturers in Japan.
Unless otherwise specified, machining tolerances conformed to Japanese Industrial Standard (JIS) B 0419-mK (comparable to ISO 2768). A surface roughness of Ra~6.3~\si{\micro\meter} was specified for exterior machined surfaces, while hermetic sealing surfaces were specified to be finished as a circumferential surface roughness of Ra~1.6~\si{\micro\meter}. All oxygen-free copper components were gold plated without a nickel underlayer (approximately 0.4~\si{\micro\meter} thick).

\begin{figure}[tb]
 \centering
\includegraphics[]{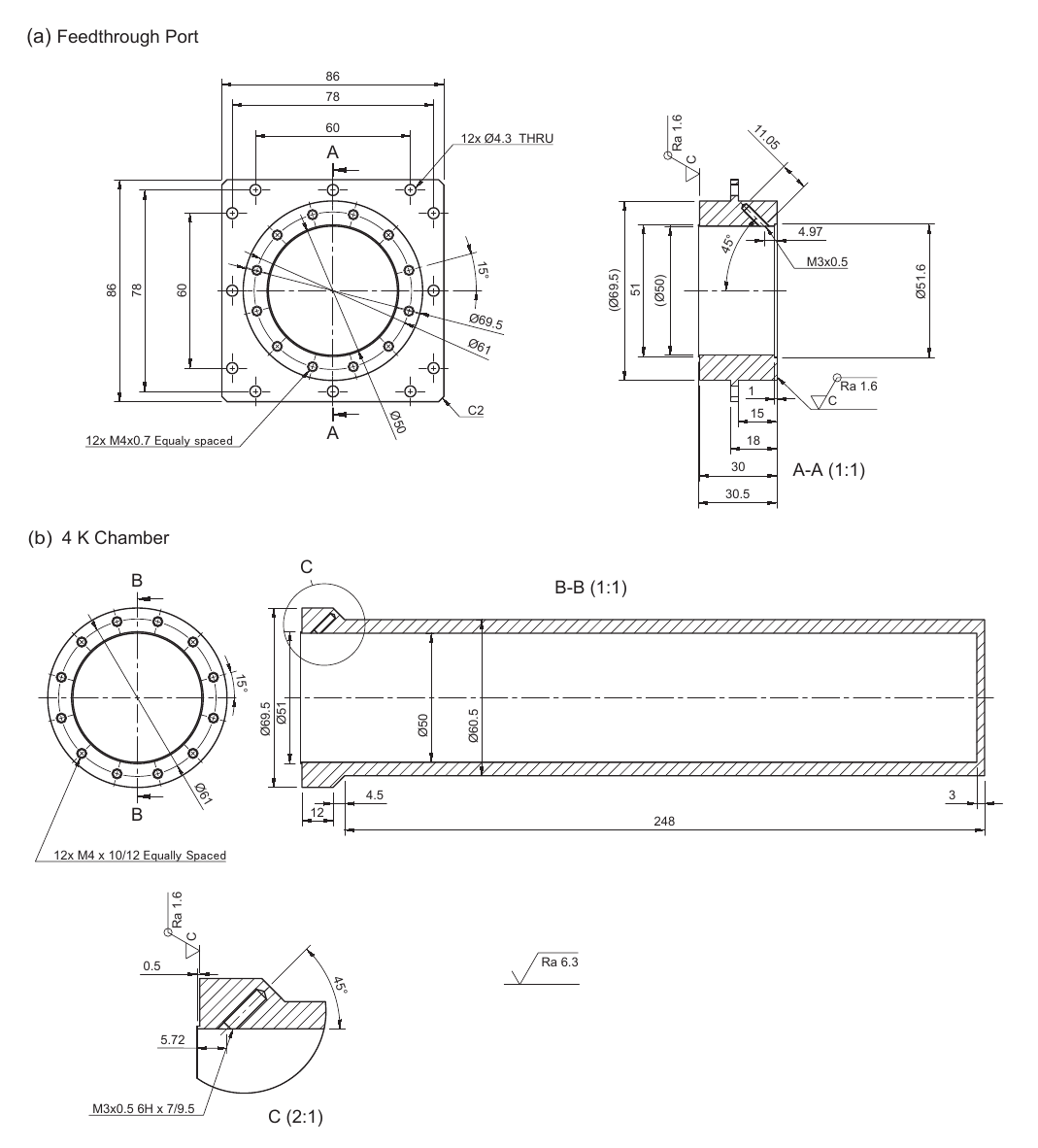}
 \caption[]{Drawings of (a) feedthrough port and (b) 4~K chamber. Dimensions are in mm.}
 \label{figS03}
\end{figure}

Figures~\ref{figS03}(a) and (b) show technical drawings of the feedthrough port and the 4~K chamber, respectively. The feedthrough port is mounted on the first stage and fixed using twelve M4 stainless-steel screws through 4.3~mm diameter through holes. Twenty M4 tapped holes arranged on a pitch-circle diameter of 61~mm are used to connect Bellows Pipe A and Bellows Pipe B.
Figure~\ref{figS03}(b) shows the 4~K chamber, which is a cylindrical chamber with an inner diameter of 50~mm and a depth of approximately 260~mm, designed to completely accommodate the approximately 200~mm-long CCC probe head. Similar tapped holes were machined on its sealing surface for connection to Bellows Pipe B. Two 45$^{\circ}$-tilted M3 blind tapped holes machined on the inner wall are used as thermal anchoring points for connection to the thermal anchors through oxygen-free copper braids terminated with crimped lugs.

\begin{figure}[tb]
 \centering
\includegraphics[]{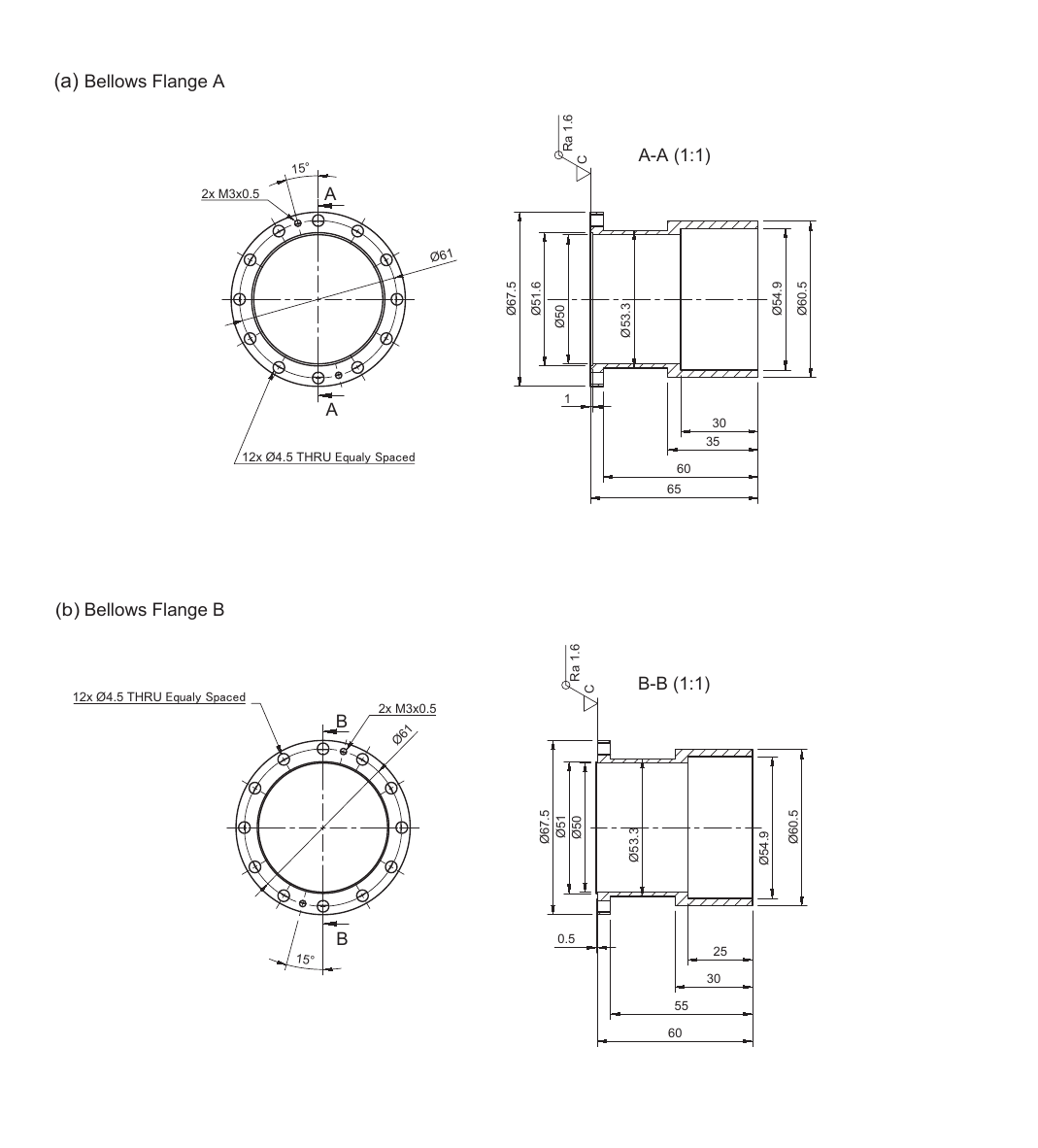}
 \caption[]{Drawings of (a) Bellows flange A and (b) B. Dimensions are in mm.}
 \label{figS04}
\end{figure}

\begin{figure}[tb]
 \centering
\includegraphics[]{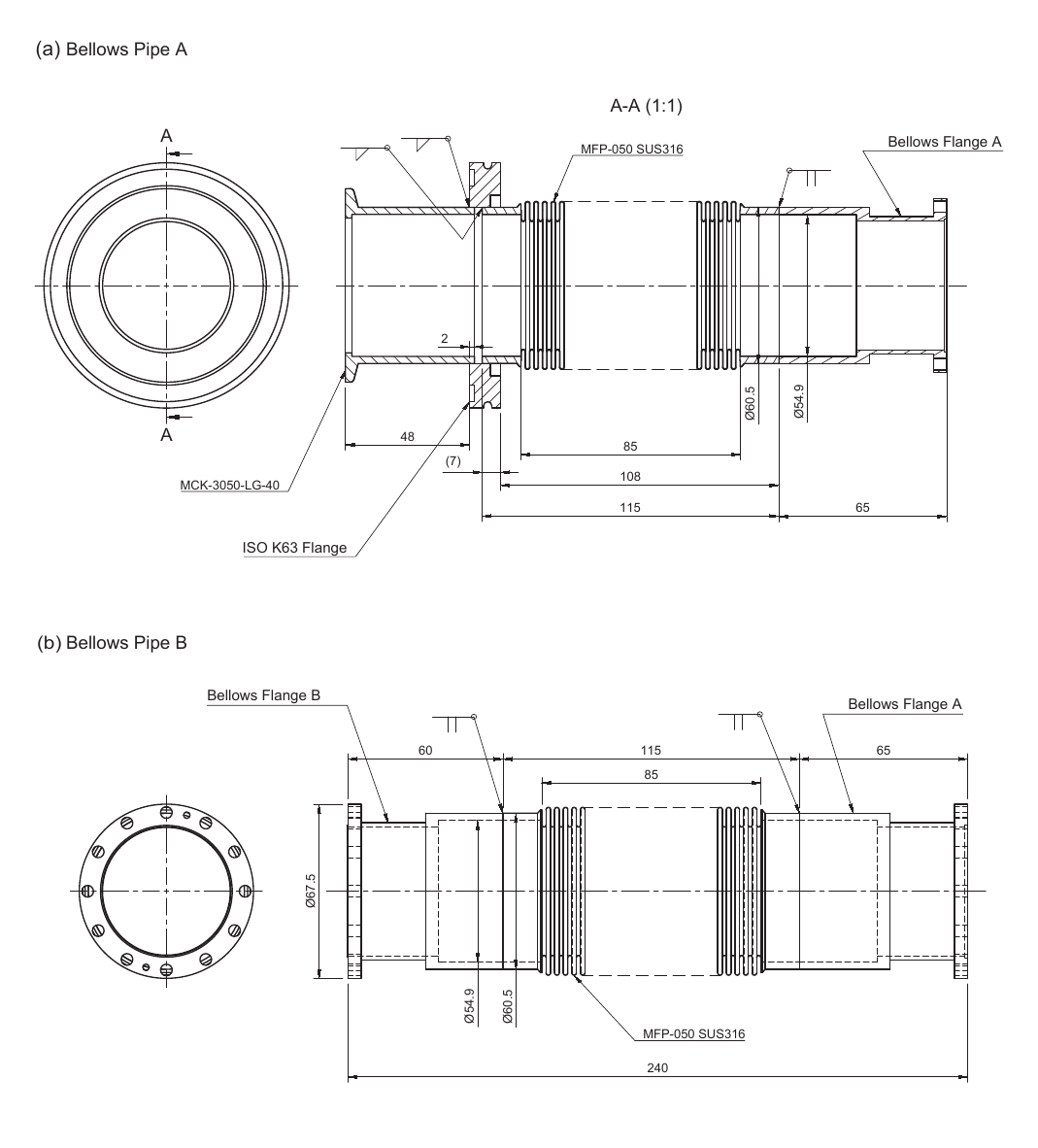}
 \caption[]{Drawings of (a) Bellows pipe A and (b) B. Dimensions are in mm.}
 \label{figS05}
\end{figure}

Bellows Pipe A and Bellows Pipe B, shown in Figs.~\ref{figS04} and \ref{figS05}, were fabricated from commercially available bellows tubing (MFP-050, MIRAPRO Co., Ltd.). The bellows section is 85~mm long and is terminated by 15~mm-long straight tube sections at both ends, resulting in a total length of 115~mm.
Bellows Pipe A was fabricated by tungsten inert gas (TIG) welding the bellows tube to commercially available KF50 and ISO-K63 vacuum flanges and the custom Bellows Flange A shown in Fig.~\ref{figS04}(a). Similarly, Bellows Pipe B was fabricated by TIG welding the same bellows tube to the custom Bellows Flanges A and B shown in Fig.~\ref{figS05}.

\begin{figure}[tb]
 \centering
\includegraphics[]{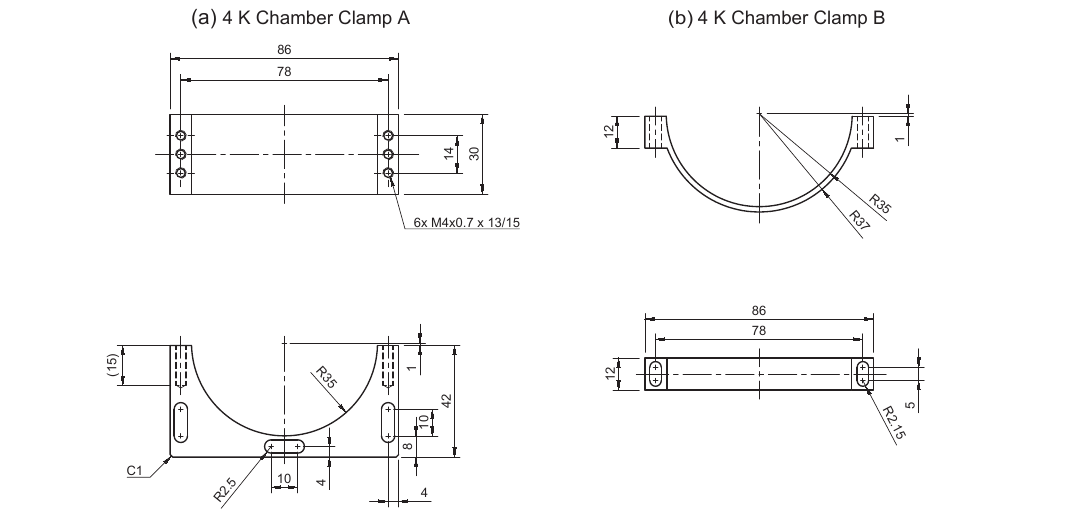}
 \caption[]{Drawings of (a) 4 K chamber clamp A and (b) B. Dimensions are in mm.}
 \label{figS06}
\end{figure}

\begin{figure}[tb]
 \centering
\includegraphics[]{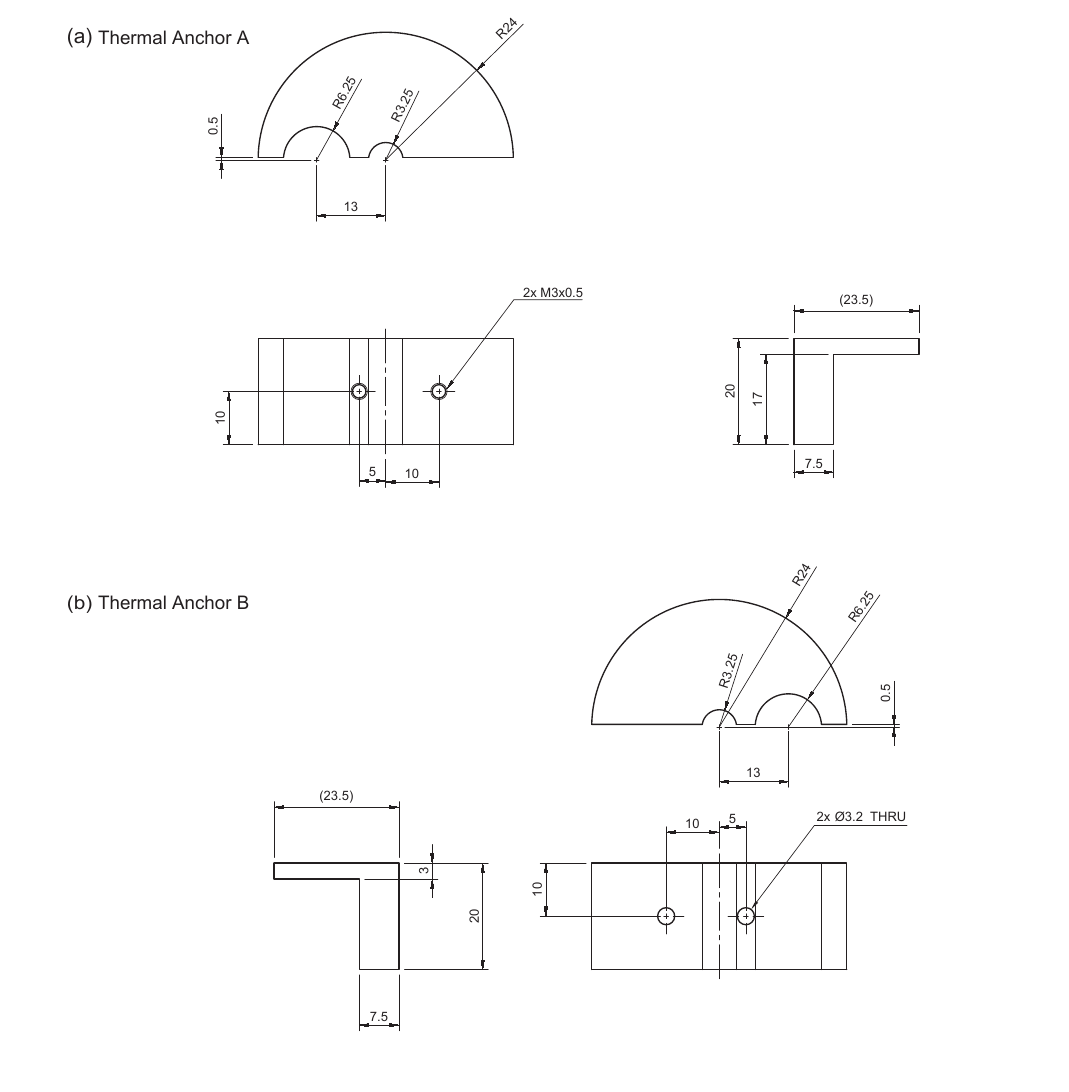}
 \caption[]{Drawings of (a) Thermal anchor A and (b) B. Dimensions are in mm.}
 \label{figS07}
\end{figure}

The 4 K Chamber Clamps A and B shown in Fig.~\ref{figS06} were used to thermally anchor the 4 K chamber with the second stage. Clamp A was attached to the second stage using three 40~mm long M4 screws, while Clamp B was fastened to Clamp A to hold the 4 K chamber in place.
A pair of Thermal Anchors A and B shown in Fig.~\ref{figS07} was used for thermal anchoring. Thermal Anchor B was attached to Thermal Anchor A using two M3 screws around the CCC support pipe. The crimped terminals of the copper braids used as internal thermal links were attached to these screws prior to assembly. The thermal anchors were fabricated as 48-mm-diameter discs, which also serve as radiation shields. 

\begin{figure}[tb]
 \centering
\includegraphics[]{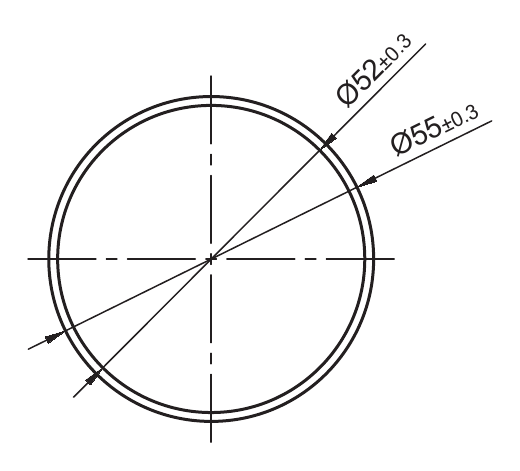}
 \caption[]{Drawing of polyimide gasket. Dimensions are in mm.}
 \label{figS08}
\end{figure}

The polyimide gasket used for hermetic sealing of low-temperature joints, shown in Fig.~\ref{figS08}, was fabricated from a 0.125~mm-thick polyimide film (DuPont 500H) by laser cutting into an O-ring shape. This gasket was used at all low-temperature joints between Bellows Pipe A and Feedthrough Port, Feedthrough Port and Bellows Pipe B, and Bellows Pipe B and the 4~K chamber. Prior to assembly, the flange surfaces were carefully cleaned with alcohol to remove dust and grease. The polyimide gasket was coated with vacuum grease and fitted around a 51~mm diameter annular protrusion (0.5~mm high) machined on the sealing surface (see Figs.~\ref{figS03}(a), (b) and \ref{figS04}). Finally, the joint was tightened uniformly using twelve M4 screws. The gasket is intended for single use and should be replaced after disassembly.

\begin{figure}[tb]
 \centering
\includegraphics[]{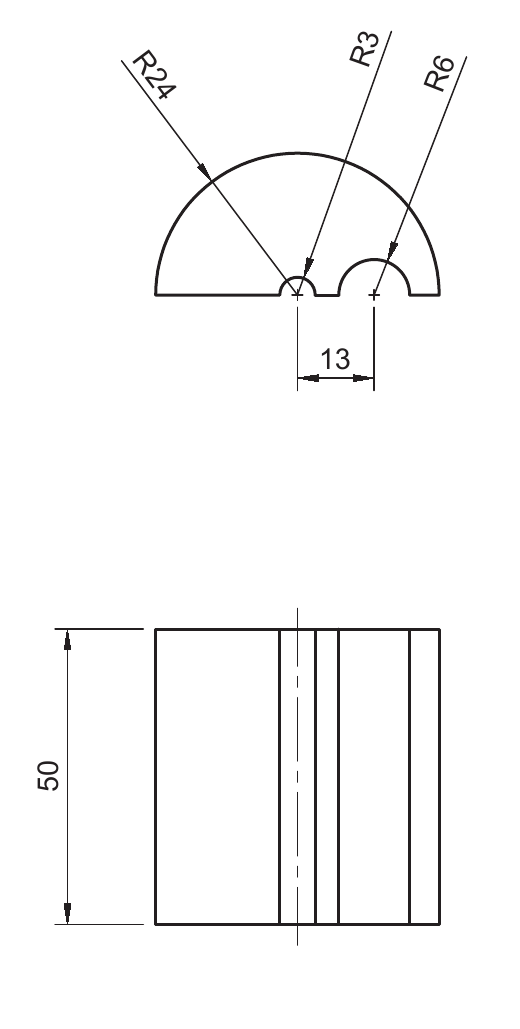}
 \caption[]{Drawing of Glasswool insulator. Dimensions are in mm.}
 \label{figS09}
\end{figure}

The glass-wool insulator shown in Fig.~\ref{figS09} was fabricated from a 50~mm-thick glass-wool board by water-jet cutting the board into a half-cylindrical shape with a diameter of 48~mm. The radial clearance between the glass-wool insulator and the inner wall of the bellows pipe (49.5~mm inner diameter) was designed to be only 1.5~mm, which is sufficiently small to suppress gas convection and acoustic noise. The board was produced by stacking and bonding two 25~mm-thick glass-wool sheets (Paramount Glass MFG. Co., Ltd.) with a density of 96~$\mathrm{kg/m^3}$. A pair of half-cylindrical glass-wool pieces was assembled around the CCC support pipe using PTFE tape.

\section{Stage temperature stability and resistance ratio measurements}

\begin{figure}[tb]
 \centering
\includegraphics[]{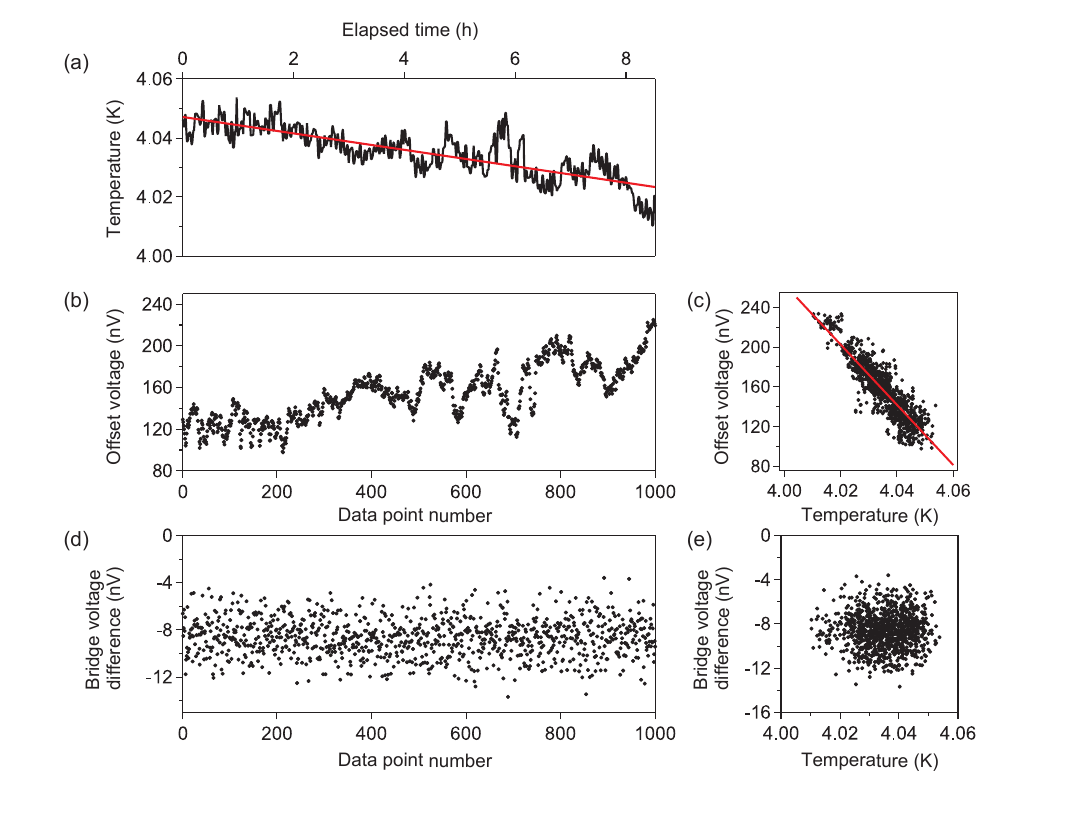}
 \caption[]{(a) Second-stage temperature as a function of elapsed time. The horizontal axis spans the same period as the 1000 consecutive measurements shown in Figs.~(b) and (d). The solid line shows the best linear fit. (b, d) Results of 1000 consecutive measurements of (b) the offset voltage and (d) the bridge voltage difference plotted as a function of data point number. (c, e) Scatter plots of (c) the offset voltage versus second-stage temperature and (e) the bridge voltage difference versus second-stage temperature.}
 \label{figS10}
\end{figure}

We investigated the stability of the refrigerator stage temperature and its impact on CCC measurements during the QHR/100~$\Omega$ resistance ratio measurements presented in Fig.~5 of the main manuscript.
The 1000 consecutive measurements shown in Fig.~5(a) are replotted in Fig.~\ref{figS10}(d). Figure~\ref{figS10}(a) shows the second-stage temperature as a function of elapsed time. The horizontal axis spans the same measurement period as that shown in Figs.~\ref{figS10}(b) and (d). The solid line represents a linear fit to the temperature data.
The temperature exhibited a linear drift of $-2.78$~mK/h during the measurement. After subtracting the linear trend, the residual temperature fluctuation had an RMS value of 4.77~mK.
To investigate the influence of the stage-temperature variations on the CCC measurements, we also plot the bridge offset voltage, defined as $(V_{+}+V_{-})/2$, in Fig.~\ref{figS10}(b), and the bridge voltage difference, defined as $(V_{+}-V_{-})$, in Fig.~\ref{figS10}(d), where $V_{+}$ and $V_{-}$ denote the measured bridge voltages at positive and negative current polarities, respectively.
The temporal variation of the offset voltage closely follows that of the stage temperature, whereas no similar behavior is observed for the bridge voltage difference. The Pearson correlation coefficient between the stage temperature and the offset voltage was $-0.8898$, indicating a strong negative correlation between these quantities. The linear fit shown in Fig.~\ref{figS10}(c) yields a temperature coefficient of $-3.03$~nV/mK for the offset voltage.
In contrast, the Pearson correlation coefficient between the stage temperature and the bridge voltage difference was only 0.012, indicating no observable linear correlation between the two quantities, consistent with the randomly distributed scatter plot shown in Fig.~\ref{figS10}(e).

The physical origin of this correlation remains unclear. One possible explanation is that temperature-dependent changes in the flux bias or current bias of the SQUID contribute to the observed offset-voltage variation.
However, a detailed investigation of the underlying mechanism is beyond the scope of the present work and remains a subject for future study.
Although a temperature-induced shift of the offset voltage was observed, this effect is effectively canceled in the bridge voltage difference in the present measurement procedure described in this work. Consequently, no observable impact of the stage-temperature fluctuations on the resistance ratio measurements was found.
We therefore conclude that the temperature stability of the refrigerator is sufficient for the precision resistance ratio measurements reported in this work.
It should be noted, however, that increasing the current reversal cycle period could introduce noise contributions that may not be canceled.

\begin{table}[t]
\centering
\caption{Components used in the helium gas chamber.}
\label{tab:parts}
\begin{tabular}{lllll}
\hline
Part name & Qty. & Material & Fabrication method & Drawing \\
\hline
Feedthrough Port & 1 & Oxygen-free Cu  & Machining & Fig.~\ref{figS03}(a) \\
4~K Chamber & 1 & Oxygen-free Cu  & Machining & Fig.~\ref{figS03}(b) \\
Bellows Flange A & 2 & 316 stainless steel & Machining & Fig.~\ref{figS04}(a) \\
Bellows Flange B & 1 & 316 stainless steel & Machining & Fig.~\ref{figS04}(b) \\
Bellows Pipe A & 1 & 316 stainless steel & TIG Welding & Fig.~\ref{figS05}(a) \\
Bellows Pipe B & 1 & 316 stainless steel & TIG Welding & Fig.~\ref{figS05}(b) \\
4~K Chamber Clamp A & 1 & Oxygen-free Cu  & Machining & Fig.~\ref{figS06}(a) \\
4~K Chamber Clamp B & 1 & Oxygen-free  Cu  & Machining & Fig.~\ref{figS06}(b) \\
Thermal Anchor A & 2 & Oxygen-free  Cu  & Machining & Fig.~\ref{figS07}(a) \\
Thermal Anchor B & 2 & Oxygen-free  Cu  & Machining & Fig.~\ref{figS07}(b) \\
Polymide Gasket & 3 & Polyimide film (t0.125 mm) & Laser cutting & Fig.~\ref{figS08} \\
Glasswool insulator & 10 & Glass wool (96~$\mathrm{kg/m^3}$) & Water jet cutting & Fig.~\ref{figS09} \\
\hline
\end{tabular}
\end{table}

